\documentclass[12pt]{article}
\usepackage[T1]{fontenc}
\newcommand{\changefont}[3]{\fontfamily{#1}\fontseries{#2}\fontshape{#3}\selectfont}
\usepackage{dcc}
\usepackage{booktabs}
\usepackage{multirow}
\usepackage{epsf,epsfig}
\usepackage[center,footnotesize]{caption}
\usepackage{url}
\usepackage{hyperref}
\usepackage[numbers]{natbib}
\usepackage{xspace}
\usepackage{graphicx}
\usepackage{wrapfig}
\usepackage{tikz}
\usetikzlibrary{matrix,arrows,decorations.pathmorphing,decorations.pathreplacing,calc, decorations.markings}

\usepackage{amsmath}
\newcommand{\method}[1]{{\sc #1}}

\newcommand{\accessop}{\method{access}}
\newcommand{\rankop}{\method{rank}}
\newcommand{\selectop}{\method{select}}

\newcommand{\algname}[1]{\mbox{#1}}
\newcommand{\SORT}{\algname{SORT}}
\newcommand{\SADA}{\algname{SADA}}
\newcommand{\GREEDY}{\algname{GREEDY}}

\newcommand{\structname}[1]{\mbox{#1}}
\newcommand{\CSA}{\structname{CSA}}
\newcommand{\CST}{\structname{CST}}
\newcommand{\RMQ}{\structname{RMQ}}
\newcommand{\BV}{\structname{BV}}
\newcommand{\SA}{\structname{SA}}
\newcommand{\WT}{\structname{WT}}

\newcommand{\inproteins}{{\sc proteins}}
\newcommand{\inenwiki}{{\sc enwiki}}
\newcommand{\inenwikisml}{{\sc enwiki-sml}}
\newcommand{\inenwikibig}{{\sc enwiki-big}}

\newcommand{\identifier}[1]{{\tt #1}}
\newcommand{\csafull}{\identifier{csa\_full}}
\newcommand{\border}{\identifier{border}}
\newcommand{\borderrank}{\identifier{border\_rank}}
\newcommand{\borderselect}{\identifier{border\_select}}
\newcommand{\rminq}{\identifier{rminq}}
\newcommand{\rmaxq}{\identifier{rmaxq}}
\newcommand{\docisa}{\identifier{doc\_isa}}
\newcommand{\wtd}{\identifier{wtd}}

\newcommand{\Order}[1]{\ensuremath{\mathcal{O}(#1)}}

\newcommand{\TEXT}{\ensuremath{\mathcal{T}}}
\newcommand{\ALPH}{\ensuremath{\Sigma}}
\newcommand{\ALPHSIZE}{\ensuremath{\sigma}}
\newcommand{\DOCS}{\ensuremath{\mathcal{C}}}
\newcommand{\QUERY}{\ensuremath{\mathcal{Q}}}
\newcommand{\DA}{\ensuremath{\mathcal{D}}}
\newcommand{\SUF}{\ensuremath{\mathcal{SA}}}
\newcommand{\spp}{\mbox{\emph{sp}}}
\newcommand{\epp}{\mbox{\emph{ep}}}

\newcommand{\gb}[1]{{\mbox{$#1$~GB}}}
\newcommand{\mb}[1]{{\mbox{$#1$~MB}}}
\newcommand{\kb}[1]{{\mbox{$#1$~kB}}}

\DeclareMathOperator{\tf}{tf}
\DeclareMathOperator{\df}{df}

\newcommand{\pizzachili}{{\sc{Pizza\&Chili}}}
\newcommand{\sdsl}{{\sc{sdsl}}}
\newcommand{\SDS}{{\sc{SDSs}}}

\newcounter{todocount}
\newcommand{\sunburstbullet}[4]{
\node[below right,inner ysep=4.1mm,align=left] (#1) at (#2.south west) {#3\\(\ensuremath{#4})};
\node[left,circle,fill=black,inner sep=0.5mm] (#1_bullet) at (#1.west) {};
}

\newcommand{\sunburstconnect}[3]{
\node[circle,fill=white,inner sep=0.6mm] (s_#3s) at (#1,#2) {};
\node[circle,fill=black,inner sep=0.5mm] (s_#3) at (#1,#2) {};
\node[circle,fill=white,inner sep=0.3mm] (m_#3s) at (medx |- l_#3) {};
\node[circle,fill=black,inner sep=0.2mm] (m_#3) at (medx |- l_#3) {};
\path[draw=white,ultra thick] (s_#3) -- (m_#3) -- (l_#3_bullet);
\path[draw=black,thick] (s_#3) -- (m_#3) -- (l_#3_bullet);
}

\newcommand{\sourcedir}{http://people.eng.unimelb.edu.au/sgog/DCC2014}
\newcommand{\dsl}[1]{%
\sourcedir/info/#1.html%
}%
\newcommand{\dsu}[2]{%
\href{\dsl{#1}}{#2}%
}%

\newcommand{\sdslgitinc}[0]{https://github.com/simongog/sdsl-lite/tree/master/include/sdsl}
\newcommand{\sdslclass}[2]{\href{\sdslgitinc/#2}{\changefont{cmtt}{m}{sl}#1}}
\newcommand{\rrrvector}{{\sdslclass{rrr\_vector}{rrr_vector.hpp}}}
\newcommand{\wthuff}{{\sdslclass{wt\_huff}{wt_huff.hpp}}}
\newcommand{\wtint}{{\sdslclass{wt\_int}{wt_int.hpp}}}
\newcommand{\bitvector}{{\sdslclass{bit\_vector}{int_vector.hpp}}}
\newcommand{\ranksupportvV}{{\sdslclass{rank\_support\_v5}{rank_support_v5.hpp}}}
\newcommand{\csasada}{{\sdslclass{csa\_sada}{csa_sada.hpp}}}
\newcommand{\csawt}{{\sdslclass{csa\_wt}{csa_wt.hpp}}}
\newcommand{\sdvector}{{\sdslclass{sd\_vector}{sd_vector.hpp}}}
\newcommand{\rmqsuccinctsct}{{\sdslclass{rmq\_succinct\_sct}{rmq_succinct_sct.hpp}}}
\newcommand{\intvector}{{\sdslclass{int\_vector}{int_vector.hpp}}}

\newcommand{\bytealphabet}{{\sdslclass{byte\_alphabet}{csa_alphabet_strategy.hpp}}}
\newcommand{\intalphabet}{{\sdslclass{int\_alphabet}{csa_alphabet_strategy.hpp}}}
\newcommand{\memorymonitor}{{\sdslclass{memory\_monitor}{memory_management.hpp}}}

\title{From Theory to Practice: Plug and Play with Succinct Data Structures}

\author{
    Simon Gog${}^1$
		{\qquad}
    Timo Beller${}^2$
		{\qquad}
    Alistair Moffat${}^1$
		{\qquad}
    Matthias Petri${}^{1}$
}

\date{November 2013}

\begin{document}
\maketitle

\footnotetext[1]{Dept.\ Computing and Information Systems,
        The University of Melbourne, Victoria 3010, Australia.}
\footnotetext[2]{Inst.\ Theoretical Computer Science, Ulm University,
D-89069, Germany.}

\vspace*{-5ex}
\begin{abstract}
Engineering efficient implementations of compact and succinct
structures is a time-consuming and challenging task, since there is
no standard library of easy-to-use, highly optimized, and composable
components.
One consequence is that measuring the practical impact of new
theoretical proposals is a difficult task, since older baseline
implementations may not rely on the same basic components, and
reimplementing from scratch can be very time-consuming.
In this paper we present a framework for experimentation with
succinct data structures, providing a large set of configurable
components, together with tests, benchmarks, and tools to analyze
resource requirements.
We demonstrate the functionality of the framework by recomposing
succinct solutions for document retrieval.
\end{abstract}

%
%
%
%
%
%
%
%
%
%
%
%
%
\pagestyle{plain}
\thispagestyle{plain}

\section{Introduction}
\label{sec-introduction}

The field of succinct data structures (\SDS) has evolved rapidly in
the last decade.
New data structures such as the FM-Index, Wavelet Tree (\WT), Range
Minimum Query Structure (\RMQ), Compressed Suffix Array (\CSA), and
Compressed Suffix Tree (\CST) have been developed, and been shown to
be remarkably versatile.
These structures provide the same functionality as the corresponding
uncompressed data structures, but do so using space which is
asymptotically close to the information-theoretic lower bound needed
to store the underlying data or objects.
Using standard models of computation, the asymptotic runtime
complexity of the operations performed by {\SDS} is also often
identical to their classical counterparts.
However, in practice, {\SDS} tend to be slower than the uncompressed
structures, due to more complex memory access patterns on bitvectors,
including non-sequential processing of unaligned bits.
That is, they come in to their own only when the data scale means
that an uncompressed structure would not fit in to main memory (or
any particular level of the memory hierarchy), but a compressed
structure would.

Accessing (\accessop), counting (\rankop), and selecting (\selectop)
bits in bitvectors ({\BV}s) are the fundamental operations from which
more intricate operations are constructed.
All three foundational operations can be supported in constant time
adding only sublinear space.
Wavelet trees build on {\BV}s and generalize the three operations to
alphabets of size $\sigma>2$, with a corresponding increase in time
to $\Order{\log\sigma}$.
A further layer up, some {\CSA}s use {\WT}s to realize their
functionality; in turn, {\CSA}s are the basis of yet more complex
structures such as {\CST}s.
Multiple alternatives exist at each level of this dependency
hierarchy.
For example, {\WT}s differ in both shape (uniform versus
Huffman-shaped) and in the choice made in the lower hierarchy levels
(whether compressed or uncompressed {\BV}s are used).
The diversity of options allows structures to be composed that have a
variety of time-space trade-offs.
In practice, many of the complex structures that have been proposed
are not yet fully implemented, since the implementations of
underlying structures are missing.
The use of non-optimized or non-composable substructures then
prevents thorough empirical investigations, and makes it difficult to
carry out impartial comparisons between complex structures.
The cost of implementing different approaches also creates a barrier
that makes it difficult for new researchers, including graduate
students, to enter the field.

As part of our investigation into {\SDS}, a library of flexible and
efficient implementations has been assembled, providing a modular
``plug and play, what you declare is 
what you get'' 
approach to
algorithm implementation.
Instrumentation and visualization tools are also included in the
library, allowing space costs to be accurately measured and depicted;
as are efficient routines for constructing (in-memory as well as
semi-external) and serializing all internal representations, allowing
files containing succinct structures to be generated and re-read.
We have also incorporated recent hardware developments, so that
built-in popcount operations are used when available, and hardware
hugepages can be enabled, to bypass address translation costs.

With this resource it is straightforward to {\emph{compose}} complex
structures; {\emph{measure}} their behavior; and {\emph{explore}} the
range of alternatives.
Having an established and robust code base also means that
experimental comparisons of new data structures and algorithms can be
made more resilient, providing greater consistency between
alternative structures, and allowing better baseline systems and
hence more reproducible evaluations.
For example, different {\CST} components might be appropriate to
different types of input sequence (integer alphabet versus
byte-based; highly-repetitive or not; and sampling rates for access
methods); with the library the right combination of components for
any given application can be readily determined.
The generality embedded in the library means that it is also useful
in other fields such as information retrieval
and bioinformatics.

We illustrate the myriad virtues of the library -- version~2 of
{\sdsl} -- through two case studies: first, a detailed recomposition
and re-evaluation of succinct document retrieval systems
(Section~\ref{sec-recomp}); and second, a detailed examination of the
construction processes used to create indexing structures
(Section~\ref{sec-constr}).
%
%
%

%
%
%
%
%
%
%
%
%
%
%
%
%
%
%
%
%
%
%
%
%
%
%
%
%
%
%
%
%
%
%
%
%
%
%
%
%
%
%
%
%
%
%
%
%
%
%
%
%
%
%
%
%
%
%
%
%
%
%
%
%
%
%
%
%
%
%
%
%
%
%
%
%
%
%
%
%
%
%
%
%
%
%

\section{Document Retrieval Recomposed}
\label{sec-recomp}

The {\emph{top-$k$ document retrieval problem}} is fundamental in
information retrieval (IR), and has become an active research topic
in the succinct data structures community, see {\citet{n14compsurv}}
for an excellent overview.
For a collection of $N$ documents $\DOCS\!=\!\{d_1,\ldots,d_N\}$ over
an alphabet $\ALPH$ of size $\ALPHSIZE$, a query {\QUERY} also a set
of strings over $\ALPH$, and a ranking function
$R:\DOCS\!\times\!\QUERY\!\rightarrow\!\mathcal{R}$, the task is to
return the $k$ documents with highest values of $R(d_i,\QUERY)$.
The simple {\emph{frequency}} version of the problem assumes that the
query consists of a single sequence (term) $q$ and that $R(d_i,q)$ is
the frequency $\tf(d_i,q)$ of $q$ in $d_i$.
The {\emph{tf-idf}} version of the problem computes
$R(d_i,\QUERY)=\tf(d_i,q)\times\log({N}/{\df(q)})$, where $\df(q)$ is
the {\emph{document frequency}} of $q$, the number of documents in
which sequence $q$ occurs.
The {\emph{tf-idf}} formulation used here and in similar studies (for
example, {\citet{sad07tcs}}) is still a simplification of the
measures used in modern retrieval systems, which incorporate factors
like document length, static document components such as pagerank,
and queries with multiple terms.

We focus here on the single term {\emph{frequency}} and single term
{\emph{tf-idf}} versions of the problem.
{\citet{sad07tcs}} devised the first succinct structure for these
problems, an approach we refer to as {\SADA}.
An alternative mechanism, {\GREEDY}, was presented by
{\citet{CUL:NAV:PUG:TUR:2010}}.
{\citeauthor{CUL:NAV:PUG:TUR:2010}}~also describe implementations of
{\SADA} and {\GREEDY}, based on components that were available in
2009.
We first briefly explain {\SADA} and {\GREEDY}, and then recompose
and reimplement them using the library,
to study the impact of state-of-the-art components.
For both solution we use the conventions established by
{\citeauthor{sad07tcs}} and {\citeauthor{CUL:NAV:PUG:TUR:2010}}: the
set of documents is concatenated to form a text {\TEXT} by appending
a sentinel symbol $\#$ to each $d_i$; then joining them all in to
one; then appending a further sentinel symbol $\$$ following $d_N$.
Both sentinels are lexicographical smaller than any symbol in
$\ALPH$, and $\$<\#$.
We use $n$ to represent $|{\TEXT}|$.

\subsection{SADA and GREEDY revisited}

The {\SADA} structure is composed of several components.
First, a {\CSA} (denoted {\csafull}) over {\TEXT} identifies, for any
pattern $p$, the range $[\spp..\epp]$ of matching suffixes in
{\TEXT}, providing the functionality of a {\SA}.
Second, a {\BV} is constructed (denoted $\border[0..(n-1)]$) with
$\border[i]\!=\!1$ iff $\TEXT[i]\!=\!\#$; supporting rank and select
structures ($\borderrank$ and $\borderselect$) are also generated.
A {\emph{document array}} $\DA[0..(n\!-\!1)]$ that maps the $i$\,th
suffix in $\TEXT$ to the document that contains it can then be
emulated using {\border} and {\csafull}, since
$\DA[i]\!=\!\borderrank(\SUF[i])$.
Third, to generate all distinct document numbers in a range
$[\spp..\epp]$, an {\RMQ} structure $\rminq$ is used.
It is built over a conceptual array $C[0..(\!n-\!1)]$, defined as
$C[i]\!=\!\max\{j\!\mid\!j\!<\!i\!\wedge\!D[j]\!=\!D[i]\}$; that is, $C[i]$ is the
index in $\DA$ of the last prior occurrence of $\DA[i]$.
The number of distinct values in $\DA[\spp..\epp]$ corresponds to
$\df(q)$.
Locations of the values are identified by computing $x=\rminq(\spp,\epp)$, the
index of the minimum element in $C[\spp..\epp]$.
A temporary {\BV} of size $N$ is used to check if $\DA[x]$ was
already retrieved; if it wasn't, the counting is continued by
recursing into $[\spp..(x\!-\!1)]$ and $[(x\!+\!1)..\epp]$.
Finally, a second {\RMQ} structure and individual {\CSA}s are built
for each document $d_i$, in order to calculate $\tf(d_i,q)$.
The top-$k$ items are then calculated by a partial sort on the
$\tf(d_i,q)$ values.
In total, {\SADA} uses
$
	|\CSA(\TEXT)|\!+\!|\BV(\TEXT)|\!+\!2|\RMQ|\!+\!%
		\sum_{i=0}^{N-1}|\CSA(d_i)|
$
bits.
Choosing an $H_k$-compressed {\CSA} for $\csafull$, a
$2n\!+\!o(n)$-bit solution for $\rminq$, and a compressed {\BV} for
$\border$~{\cite{OKA:SAD:2007}}, results in a total bit count bounded
above by
$
	n H_k(\TEXT)\!+\!\sum_{i=0}^{N-1}|\CSA(d_i)|%
		+\!4n\!+\!o(n)\!+N(2\!+\!\log (n/N)).
$
The space for the document {\CSA}s was deliberately not substituted
by the cost of a concrete solution, for two reasons: (1) each {\CSA}
has an alphabet dependent overhead, which dominates the space for
small documents; and (2) only the inverse {\SA} functionality is
used.
Thus, in the recomposition, we opt for a {\emph{bit-compressed}}%
\footnote{%
That is, each item is encoded as a binary value in $\lceil\log|d_i|\rceil$ bits.
}
version of
the inverse {\SA}, with almost no overhead per document, and
benefiting from constant access time.

The second solution -- {\GREEDY} -- also uses {\csafull} to translate
patterns $p$ to ranges $[\spp..\epp]$.
In this solution the document array $\DA$ is explicitly represented
by a {\WT}, denoted {\wtd}.
The total size is then
$
	|\CSA(\TEXT)|+|\WT(\DA)|=nH_k(\TEXT)+nH_0(\DA),
$
using compressed {\BV}s for
{\wtd}~\cite{RAM:RAM:RAO:2002}.
{\citeauthor{CUL:NAV:PUG:TUR:2010}}~use these structures to solve the
frequency variant of the top-$k$ problem.
In the first step the pattern is translated to a range in $\DA$.
A priority queue is then used to store pending nodes in the expansion
in $\WT$ of the range $[\spp..\epp]$.
At each step the largest node is extracted, and, if it is not a leaf
node, its two children inserted.
This process is iterated until $k$ leaves -- corresponding to the
top-$k$ frequent documents -- have emerged.
The range size of a leaf corresponds to the term frequency $\tf(d_i,q)$.
In contrast to {\SADA}, not all documents have to be listed in order
to calculate the top-$k$.
However, {\GREEDY} does more work per document, since {\wtd} has a
height of $\log N$.

\subsection{Experimental setup}

We adopt the experimental setup employed by
{\citet{CUL:NAV:PUG:TUR:2010}}, reusing the
{\inproteins} file and adding the {\inenwiki}
file.\footnote{We did not use the Wall
Street Journal file, since licensing issues meant that we could not
make it available for download. Our full setup is
available at \href{http://go.unimelb.edu.au/w68n}{http://go.unimelb.edu.au/w68n}.
 }
The {\inenwiki} datafiles come in four versions:
parsed as either characters or words; and either a small prefix, or
the whole collection, see Table~\ref{tbl-collections}.
The character version was generated by removing markup from a
wikipedia dump file; the word-based version by then applying
the parser of the Stanford Natural Language Group.

We generated patterns of different lengths, again following the lead
of {\citet{CUL:NAV:PUG:TUR:2010}}, with $200$ patterns of each
length; reported query times are averages over sets of $200$
patterns.
All experiments were run on a server equipped with {\gb{144}} of RAM
and two Intel Xeon E5640 processors each with a {\mb{12}} L3 cache.
The experimental code is available within the benchmark suite of the library,
and all used library classes
(printed in fixed italic font) are linked to their
definitions.

\begin{table}
\centering
{\small
\begin{tabular}{@{}l@{}l@{}r*{5}{@{}l@{\hspace{1.8ex}}r}@{}}
\toprule
Collection && $n$ && $N$ && $\Sigma_{i=0}^{N-1}\frac{|d_i|}{N}$&&$\ALPHSIZE$&& $|\TEXT|$ in MB && $\approx H_k(\TEXT)$\\[0ex]
\cmidrule{3-3}
\cmidrule{5-5}
\cmidrule{7-7}
\cmidrule{9-9}
\cmidrule{11-11}
\cmidrule{13-13}
\multicolumn{2}{r}{\textit{character alphabet}} \\
{\sc  proteins }&& 58,959,815 && 143,244          && 412    && 40&&    56 && 0.90 \\
{\sc  enwiki-sml }&& 68,210,334 && 4,390          && 15,538 &&206&&    65 && 2.01 \\
{\sc  enwiki-big }&& 8,945,231,276 && 3,903,703   && 2,291  &&211&& 8,535 && 2.02 \\

\multicolumn{2}{r}{\textit{word alphabet}} \\
{\sc  enwiki-sml}&& 12,741,343    && 4,390     && 2,902 &&   281,577&& 29 && 5.03 \\
{\sc  enwiki-big}&& 1,690,724,944 && 3,903,703 &&  433 &&  8,289,354&& 4,646 && 4.45 \\
\bottomrule
\end{tabular}
}
\caption{Collection statistics: number of characters/words, number of
documents, average document length, total collection size, and
approximate $H_k$ (determined using {\tt xz --best}) in bits per
character/word. The character based collections use one byte
per symbol, while $\lceil\log\ALPHSIZE\rceil$ bits
are used the word based case.
\label{tbl-collections}}
\end{table}

\subsection{Plug and play}

We start by composing an instance of {\GREEDY}.
For {\tt{csa\_full}}, {\sdsl} provides several {\CSA} types.
We opt for \csawt, which is based on a {\WT}.
Choosing a Huffman-shaped {\WT} \cite{MAK:NAV:2005} (\wthuff) and parameterizing it
with a suitable {\BV} (\rrrvector) results in an
$H_k$-compressed {\CSA}.
The \rrrvector\ implements the on-the-fly decoding recently
described by {\citet{NAV:PRO:2012}}, which provides low redundancy.
Finally, we minimize the space of the {\CSA} by sampling (inverse)
{\SA} values only every millionth position.
This does not affect the runtime of {\GREEDY}, since it does not
require SA access.
For {\wtd} we choose the alphabet-friendly {\WT} class {\wtint}
and parameterize it with a fast uncompressed {\BV}
(\bitvector) and small overhead rank structure
(\ranksupportvV).
No select functionality is required in {\GREEDY}.

We use the same toolbox to assemble a space- and time-efficient
version of {\SADA}.
The full {\CSA} in {\SADA} has to provide fast element access;
``plug-and-play'' exploration with different {\CSA}s showed
that {\csasada} {\cite{SAD:2003}} is the preferred choice in this situation.
The suffix sampling rate is set to $32$.
Similar exploration revealed that a text-order sampling strategy
yields a more attractive time-space trade-off than suffix-order
sampling, provided the Elias-Fano compressed {\BV} \cite{OKA:SAD:2007}
({\sdvector}) is used to mark the sampled suffixes.
For components {\tt{rminq}} and {\tt{rmaxq}}
we select the 
range min-max-tree based
{\RMQ} structure (\rmqsuccinctsct).
The inverse {\SA}s of the documents are represented using
bit-compressed vectors (\intvector), and the array of vectors
is denoted by {\docisa}.

Finally, we compose word alphabet versions of {\GREEDY} and {\SADA}
by replacing the character alphabet strategy (\bytealphabet)
in
the definition of
{\tt{csa\_full}} by the word alphabet strategy
({\intalphabet}).

\begin{table}
\small
\centering
\begin{tabular}{@{}l@{\hspace{0.2ex}}*{2}{rr@{\hspace{1ex}}l}c*{2}{rr@{\hspace{1ex}}l}@{}}
\toprule
\multirow{2}{*}{Collection}
&\ &\multicolumn{5}{c}{character alphabet}                           &  &\multicolumn{5}{c}{word alphabet} \\
\cmidrule{3-7}\cmidrule{9-14}
&\ &\multicolumn{2}{c}{\GREEDY}&\ &\multicolumn{2}{c}{\SADA}   &  &\multicolumn{2}{c}{\GREEDY}&\ &\multicolumn{2}{c}{\SADA}&  \\[1ex]
{\sc  proteins}&& \dsu{PROTEINS.GREEDY.byte}{162} & (2.87)&&
                  \dsu{PROTEINS.SADA.byte}{136} & (2.42) &  
                  &\multicolumn{5}{c}{\textit{-- no word parsing --}} & \\
{\sc  enwiki-sml}&& \dsu{ENWIKISML.GREEDY.byte}{130} & (2.01)&&
                    \dsu{ENWIKISML.SADA.byte}{204} & (3.13)  &  & 
                    \dsu{ENWIKISMLINT.GREEDYINT.int}{38} & (1.32)&& 
                    \dsu{ENWIKISMLINT.SADAINT.int}{50} & (1.72) &  \\
{\sc  enwiki-big}&& \dsu{ENWIKIBIG.GREEDY.byte}{27,043} & (3.17)&&
                    \dsu{ENWIKIBIG.SADA.byte}{24,404} & (2.86) & & 
                    \dsu{ENWIKIBIGINT.GREEDYINT.int}{6,786} & (1.46)&&
                    \dsu{ENWIKIBIGINT.SADAINT.int}{5,703} & (1.23) &  \\
\bottomrule
\end{tabular}
\caption{Sizes of indexes, in MB and as a
multiple of the collection.
\label{tbl-sizes}}
\end{table}

\paragraph{Memory Usage.}

\begin{wrapfigure}[16]{right}{7.5cm}
\centering
\vspace*{-6ex}
\begin{tikzpicture}
\def\sunheight{6.3cm}

\node[overlay,right,inner sep=0,above right] (sunburst) at (0cm,0cm) {\scalebox{0.40}{\href{\dsl{ENWIKIBIG.SADA.byte}}{\includegraphics{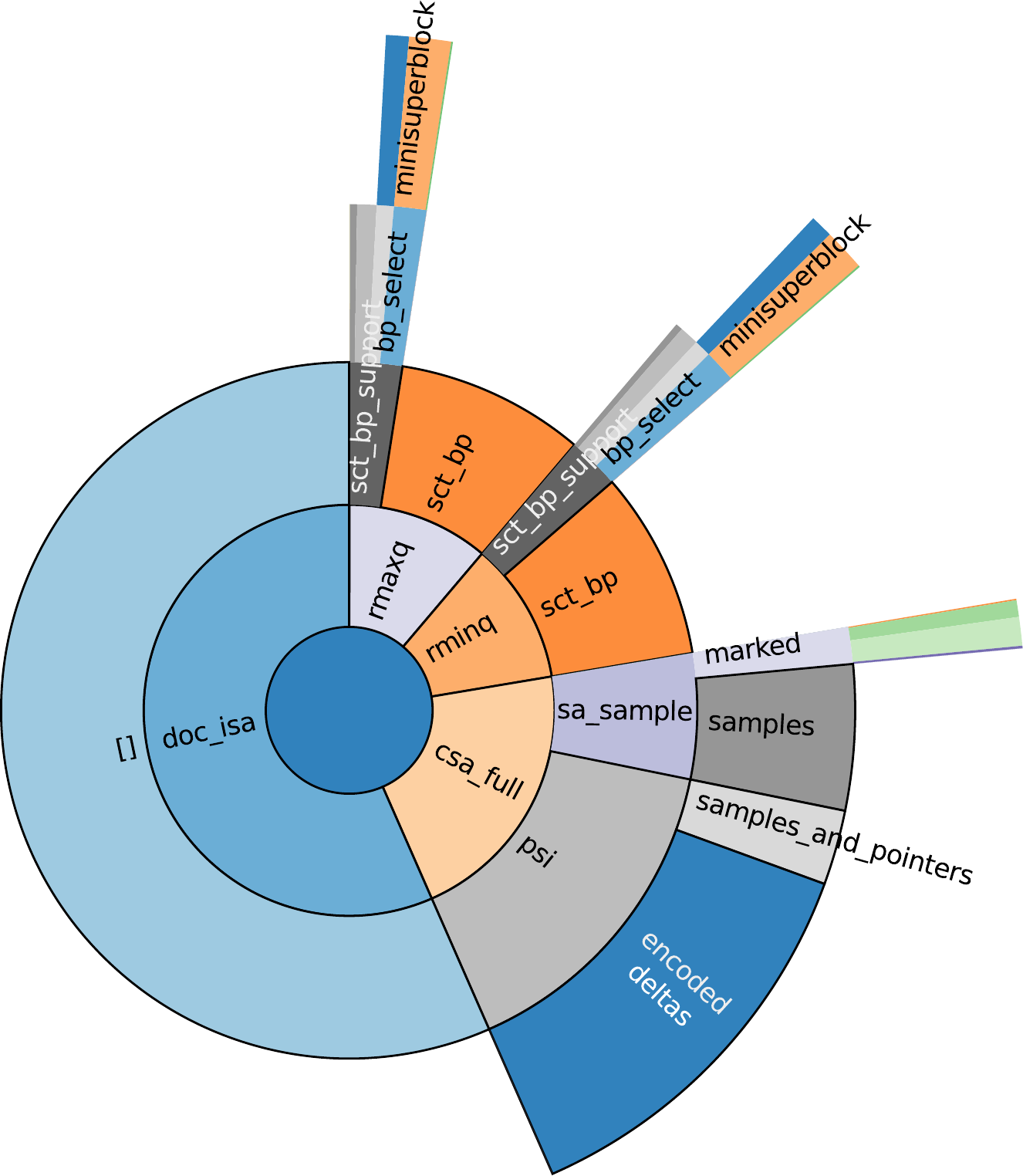}}}};

\draw[white] (0cm,0cm) rectangle(7.0cm,\sunheight);

\coordinate (legend) at (5cm, 6.1cm);
\coordinate (medx) at (3.3cm, 7cm);
\sunburstbullet{l_docisa}{legend}{\docisa}{56.6\%}
\sunburstbullet{l_rmaxq}{l_docisa}{\rmaxq}{11.2\%}
\sunburstbullet{l_rminq}{l_rmaxq}{\rminq}{11.2\%}
\sunburstbullet{l_fullcsa}{l_rminq}{\csafull}{21.0\%}

\sunburstconnect{1.10cm}{2.50cm}{docisa}
\sunburstconnect{2.25cm}{3.10cm}{rmaxq}
\sunburstconnect{2.50cm}{2.65cm}{rminq}
\sunburstconnect{2.45cm}{2.00cm}{fullcsa}

\end{tikzpicture}

\vspace*{-2ex}
\caption{Sunburst visualization of the memory
         usage of the character-based {\SADA}
         on file {\inenwikibig}. A dynamic
         version is available at
         \href{http://go.unimelb.edu.au/ba8n}{http://go.unimelb.edu.au/ba8n}.
\label{fig-sunburst}}
\end{wrapfigure}
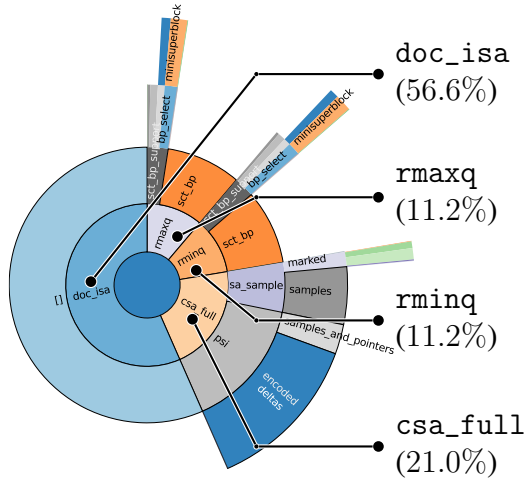

Table~\ref{tbl-sizes} summarizes the space usage of the composed
structures.
They take considerably less space than those reported by
{\citeauthor{CUL:NAV:PUG:TUR:2010}}; for example, on {\inproteins},
their {\SADA} is $6.4$ times larger than ours,
and their {\GREEDY} is $1.3$ times larger.
Also note that when the documents are short, our {\SADA} is smaller
than {\GREEDY}.
These space reductions result from the use of better-engineered
components; the only algorithmic change made was the use of
bit-compressed inverse {\SA}s, instead of high-overhead {\CSA}s.

Figure~\ref{fig-sunburst} depicts a space visualization of the type
that can be generated for any {\sdsl}
object.
It reveals that {\tt{doc\_isa}} still takes over half the space; that
{\tt{rmaxq}} and {\tt{rminq}} are close to the optimal $2n$ bits
($2.6n$ bits); and that the {\CSA} takes $5.3$ bits per character.
The latter differs from the optimal $H_k=2.02$ bits reported
in Table~\ref{tbl-sizes}, as samples were added for fast SA access,
which account for $2.4$ of the $5.3$
bits.
The largest component of {\GREEDY} (not shown here) is the document
array $\DA$, which requires $n\log N$ bits plus the overhead of the
rank structure, around $92.2\%$ of the total space for
{\inenwikibig}.
Plugging in \rrrvector\ results in $H_0$-compression of \wtd\ and reduces
the overall size to {\dsu{ENWIKIBIG.GREEDY-RRR.byte}{\mb{25,\!320}}},
while the query time is slowed down by a factor of between $2$ and $4$.
Note that the word versions are smaller than the character-based
ones, because $n$ is smaller.
{\SADA} also benefits from smaller average document lengths.
Overall the word indexes are smaller than the character-based
indexes, and at around $2/3$ of the original text size, are
comparable to compressed positional inverted files.

\paragraph{Runtime.}

\begin{figure}[t]
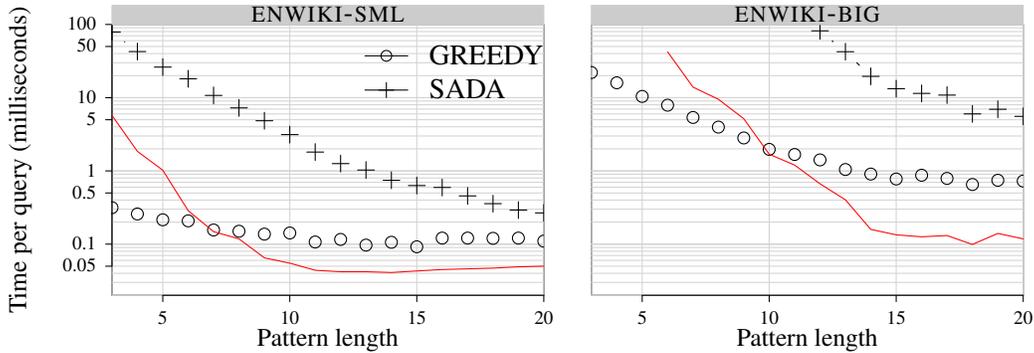

\centering
%
%

\caption{Average time for top-$10$ {\emph{frequency}} queries on the
character-based collections.
Results are omitted for pattern length $\ell$ if any single pattern
of length $\ell$ took more than $5$ seconds. The solid red line
corresponds to a baseline implementation which is explained in the text.
\label{fig-runtime}}
\end{figure}%

\begin{figure}[t]
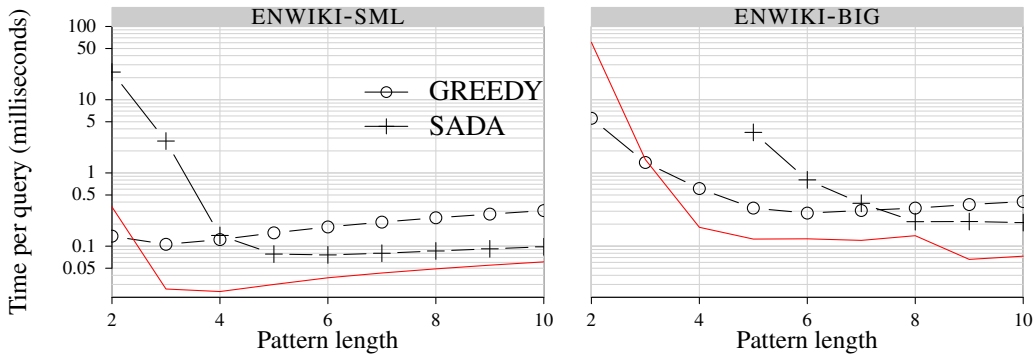

\centering
%
%
%
\caption{Average time for top-$10$ {\emph{frequency}} queries on the
word-based collections.
\label{fig-runtime_int}}
\end{figure}

Query times are depicted in Figures~\ref{fig-runtime}
and~\ref{fig-runtime_int}.
We make several observations.
First, the runtime of both solutions depend on the collection size,
since any given pattern occurs more often in a larger collection.
This results in larger ranges, which are especially bad for {\SADA},
since it processes all distinct documents in the range, even when
computing top-$10$ queries.
(This same behavior means that {\SADA} can compute complex
{\emph{tf-idf}} queries in very similar times.)
{\GREEDY} is also dependent on size of the $[\spp..\epp]$ range,
requiring two orders of magnitudes more time on {\inenwikibig} than
on {\inenwikisml}, matching the difference in their sizes.
Second, for long patterns ($>$$15$ characters) {\SADA} is now only
one order of magnitude slower than {\GREEDY}, in contrast to two
orders reported by {\citet{CUL:NAV:PUG:TUR:2010}}.
This is due to the faster extraction of inverse SA values and the
use of a $\Psi$-based {\CSA} instead of a {\WT} based one.
For word indexes {\SADA} now outperforms {\GREEDY} for long queries,
where the result range is very small.
In this cases, the pattern matching becomes the dominating cost.
In {\SADA}, we used \csasada\ which implements backward search
by binary searches on $\Psi$ using $\Order{m\log n}$ time, while
{\csawt} in {\GREEDY} performs $\Order{mH_0(\TEXT)}$ non-local rank
operations on the compressed BV.
The {\GREEDY} approach could be made significantly faster if the {\rrrvector} was
replaced by an uncompressed BV, but the space would then become
much greater than {\SADA}.

We also compare our results to a simple baseline called {\SORT}.
Again, {\csawt} is used as the {\CSA}, but now the document array \DA\ is
stored as a bit-compressed vector of $n\lceil\log N\rceil$ bits.
After identifying the $[\spp..\epp]$ interval, the entries of
$\DA[\spp..\epp]$ are copied and sorted to generate $(d_i,\tf)$
pairs; the standard C++ partial sort is then used to retrieve the
top-$10$.
The red lines in Figures~\ref{fig-runtime} and~\ref{fig-runtime_int}
show query times.
The sequential or local processing of {\SORT} is always superior to {\SADA},
which, despite careful implementation, suffers from the non-locality
of the range minimum/maximum queries.
Moreover, {\GREEDY} only dominates {\SORT} in cases involving very
wide intervals, emphasizing one of the key messages of this article
-- that careful implementations and large test instances are required
to properly show the usefulness of advanced succinct data structures.

%
%
%

%
%
%

%
%

%
%

%
%
%
%
%
%
%
%
%
%
%
%

%
%
%
%
%

%
%
%

%
%

%
%
%

%
%
%
%
%

%


\section{Efficient Construction of Complex Structures}
\label{sec-constr}

Constructing {\SDS} over small data sets -- up to hundreds of
megabytes -- is not a challenge from an engineering perspective,
since commodity hardware supports memory-space many times larger than
this.
However, {\SDS} are explicitly intended to replace traditional data
structures in resource-constrained environments; which means they
are most applicable when the data is too large for uncompressed
structures to be used, and hence that construction is also a critical
issue.
As well, complex structures composed of multiple sub-structures often
contain dependencies between sub-structures which further complicate
the construction process.
For example, to construct a {\CST}, a {\CSA} is required; and to
construct a {\CSA} fast, usually an uncompressed {\SA} is needed.
Under memory constrains, it is not possible to hold all of this
structures in RAM concurrently.
To alleviate this problem the library combines semi-external
construction algorithms which stream data which can be processed
sequentially to and from disk.
To facilitate this, the library provides serialization and save/load
functionality for all substructures.
Finally, as already noted, the library includes memory visualization
techniques, which analyze space utilization during run-time and
construction.

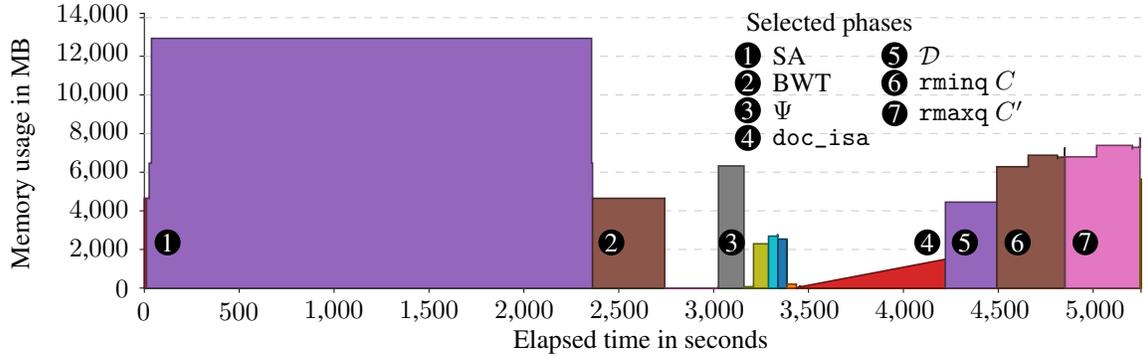
\begin{figure}[t]
\centering
%
\tikzstyle{st_ball} = [overlay,circle,font=\footnotesize,text=white,fill=black,inner sep=0.18mm]

\begin{tikzpicture}
\node[draw=white,inner sep=.3cm] at (0cm,0cm) {};
\node[draw=white,right] (img) at (1cm,0cm) {\scalebox{0.37}{\href{\sourcedir/fig-sada-int-construction.html}{\includegraphics{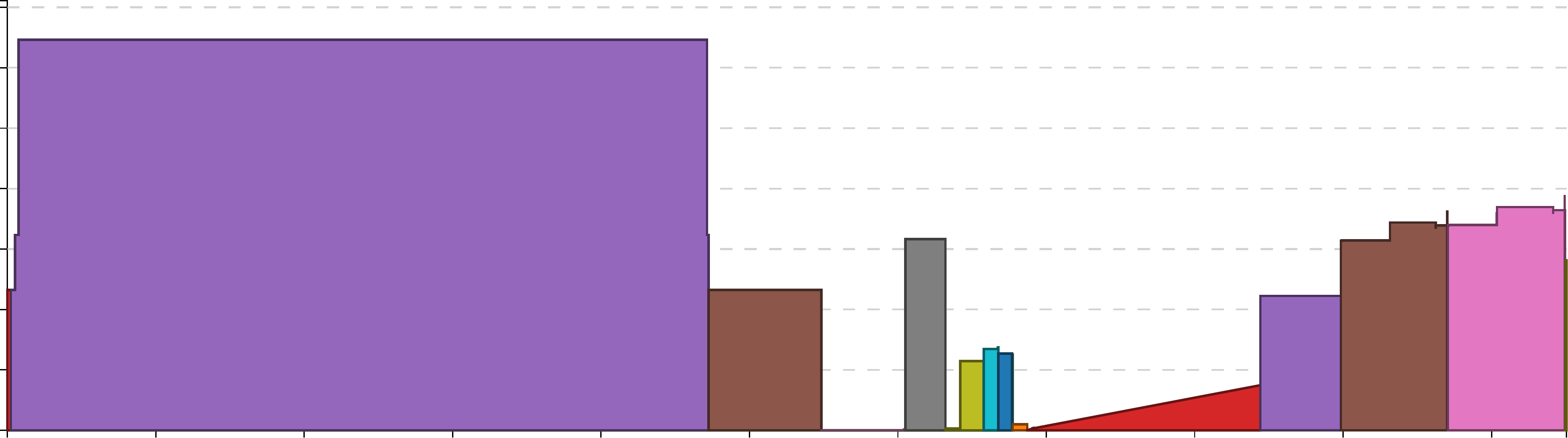}}}};
\foreach \x in {0,500,...,5000}{
    \pgfmathparse{0.2+(2.53*\x/1000)}\let\xpos\pgfmathresult
    \node[overlay, font=\footnotesize] at ($(img.south west)+(\xpos cm, -0.1cm)$) {%
        \pgfkeys{/pgf/number format/.cd,%
            fixed,%
            fixed zerofill,%
            precision=0,%
            set thousands separator={\mbox{,}}}%
        \pgfmathprintnumber{\x}%
    };
}
\foreach \iy in {0,2,...,14}{
    \pgfmathparse{0.2+(0.26*\iy)}\let\ypos\pgfmathresult
    \node[overlay, font=\footnotesize,left] at ($(img.south west)+(0.15 cm, \ypos cm)$) {
        \pgfmathparse{int(\iy*1000)}\let\yy\pgfmathresult
        \pgfkeys{/pgf/number format/.cd,%
            fixed,%
            fixed zerofill,%
            precision=0,%
            set thousands separator={\mbox{,}}}%
        \pgfmathprintnumber{\yy}%
    };
}
\node[font=\footnotesize,yshift=-0.5cm] at (img.south) {Elapsed time in seconds};
\node[font=\footnotesize,xshift=-1.4cm] at (img.west) {\rotatebox{90}{Memory usage in MB}};
\node[st_ball,xshift=-13.1cm,yshift=-3.2cm] at (img.north east) {1};
\node[st_ball,xshift=-7.2cm,yshift=-3.2cm] at (img.north east) {2};
\node[st_ball,xshift=-5.6cm,yshift=-3.2cm] at (img.north east) {3};
\node[st_ball,xshift=-3.0cm,yshift=-3.2cm] at (img.north east) {4};
\node[st_ball,xshift=-2.5cm,yshift=-3.2cm] at (img.north east) {5};
\node[st_ball,xshift=-1.8cm,yshift=-3.2cm] at (img.north east) {6};
\node[st_ball,xshift=-0.9cm,yshift=-3.2cm] at (img.north east) {7};
\node[overlay,font=\footnotesize,xshift=-4.3cm,left,below] (title) at (img.north east) {Selected phases};
\node[overlay,font=\footnotesize,below,right,yshift=-0.1cm, xshift=0.35cm] (leg1) at (title.south west) {SA};
\node[overlay,font=\footnotesize,below,right,yshift=-0.1cm] (leg2) at (leg1.south west) {BWT};
\node[overlay,font=\footnotesize,below,right,yshift=-0.1cm] (leg3) at (leg2.south west) {$\Psi$};
\node[overlay,font=\footnotesize,below,right,yshift=-0.1cm] (leg4) at (leg3.south west) {{\tt doc\_isa}};

\node[overlay,font=\footnotesize,below,right,yshift=-0.1cm,xshift=2.3cm] (leg5) at (title.south west) {\DA};
\node[overlay,font=\footnotesize,below,right,yshift=-0.1cm] (leg6) at (leg5.south west) {{\tt rminq} $C$};
\node[overlay,font=\footnotesize,below,right,yshift=-0.1cm] (leg7) at (leg6.south west) {{\tt rmaxq} $C'$};

\foreach \p in {1,2,3,4,5,6,7}
    \node[st_ball,left] at (leg\p.west) {\p};

\end{tikzpicture}
\caption{A memory-time graph for
construction of {\SADA} over the word-based
sequence {\inenwikibig} (\gb{4.6}).
The \memorymonitor\ generated version is
available at
 \href{http://go.unimelb.edu.au/7a8n}{http://go.unimelb.edu.au/7a8n}.
\label{sada-construct}}
\end{figure}

To demonstrate the complexities of the construction process in more
detail, we examine the resource utilization during the construction
of word-based {\SADA}.
Figure~\ref{sada-construct} shows the resource consumption for the
{\gb{4.6}} {\inenwikibig} collection.
In total, the construction process took $5{,}250$ seconds and had a
peak requirement of {\gb{13}}.
This corresponds to a throughput of {\mb{0.88}} per second. In monetary terms,
the \SADA\ index for {\inenwikibig} can be built for less than one dollar on the Amazon
Cloud.\footnote{In November 2013 a ``High-Memory Extra Large''
instance with {\gb{17}} RAM costs USD$\$0.41$ per hour.}
The majority of the time ($65\%$) was spent constructing the {\CSA}.
First, the plain {\SA} is constructed -- phase~$1$ in
Figure~\ref{sada-construct} -- using the algorithm of
{\citet{LAR:SAD:2007}}.
The algorithm uses twice the memory space required by the resulting
bit-compressed suffix array, accounting for the peak memory usage
(\gb{13}) of the complete process.
After construction, the {\SA} is serialized to disk to construct the
Burrows-Wheeler Transform (BWT) (phase~$2$).
Only {\TEXT} is kept in memory, as it is the only place where random
access is required, and the BWT sequence can be written to disk as it
is formed.
This semi-external construction process of the BWT requires $380$
seconds, or $7\%$ of the total time.
The remainder of the {\CSA} is constructed in $358$ seconds.
This includes constructing $\Psi$ mapping (shown as phase~$3$)
as well
as sampling the {\SA}.
The next major construction step, marked as phase $4$ in the figure,
constructs the $N$ individual inverse {\SA}s (\docisa).
Here for each document a
{\SA} is constructed,
inverted, bit-compressed, and added to {\docisa}.
This process requires $785$ seconds, or around $200$ microseconds per
document.
Next, phase~$5$ constructs the document array $\DA$ by
streaming the full {\SA} from disk and performing $n$ {\rankop}
operations
on {\border}.
Creating the complete array requires $271$ seconds or $160$
nanoseconds per $\DA[i]$ value.
This includes streaming the {\SA} from disk, performing the {\rankop}
on {\border} and storing the resulting $\DA[i]$ value.
Finally, {\rminq} (phase~$6$) and {\rmaxq} (phase~$7$) are
created, including computing temporary $C$ and $C'$ arrays from
$\DA$.
Creating {\rminq} requires $360$ seconds; computing {\rmaxq} a
further $396$ seconds.
Half of that time is spent creating $C$ and $C'$; the balance on
constructing the actual {\RMQ} structures.

Semi-external construction is an important tool to minimize
the resource consumption in each phase of the building process of
SADA.
Keeping all constructed components in memory would significantly
increase the memory overhead during the later stages of the
construction process.
Thus, careful engineering of the construction phase of each
individual component of a complex structure is important so that the
overall resource requirements of the entire construction process is
minimized.
In our case, {\SADA} is now so efficient that we can build the
structure for the word tokenized GOV2 collection, used in the TREC
Terabyte Track, on our experimental machine.

It is important to monitor the construction costs, as the
modularity of {\SDS} can lead to unintended resource consumption.
For example, in an initial version of our {\SADA} implementation, the
{\CSA} was not serialized to disk, but kept in
memory.
Visualization of the overall construction cost of {\SADA} highlighted
this inefficiency and allowed it to be rectified.
Similarly, when reviewing the character-based {\csawt}, it became
apparent that optimizing {\WT} construction will not significantly
improve the overall process, as that phase only accounts for around
$4\%$ of the total cost.
An automatic resource tracker (\memorymonitor) facilitates
visualizing the space and time consumption of algorithms in order to provide
such insights.

\begin{wraptable}[12]{right}{5.2cm}%
\vspace*{-5.0ex}
\centering
\begin{tabular}{@{}r@{}cr@{}cr@{}}
\toprule
Prefix && \multicolumn{3}{c@{}}{Page size} \\ \cmidrule{3-5}
(MB)  && \multicolumn{1}{c}{\kb{4}} && \multicolumn{1}{c@{}}{\gb{1}} \\ \cmidrule{1-1}\cmidrule{3-3}\cmidrule{5-5}
10     &&     2  &&      2 \\
100    &&    29  &&     25 \\
1,000  &&   379  &&    307 \\
5,000  &&  2,524 &&  1,877 \\
8,535  &&  5,282 &&  4,482 \\
\bottomrule
\end{tabular}
\vspace*{-0.5ex}
\caption{Construction times (seconds) of {\csawt}
for prefixes of the character-based {\inenwikibig}
file.
\label{tbl-csa-construction}}
\end{wraptable}%

Query times for {\SDS} can be markedly
improved if {\gb{1}} pages (hugepages) are used to address memory,
rather than the standard {\kb{4}} pages, since address translation
becomes a bottleneck when pages are small~\citep{gp13spe}.
The {\sdsl} now includes memory management facilities that allow the use
of hugepages during construction as well.
We investigate the effect of hugepages on construction time by
building {\csawt}\,s of increasing size for prefixes of the character-based
{\inenwikibig} collection
(Table~\ref{tbl-csa-construction}).
For small file sizes, hugepages have a modest effect on construction
time, and the {\mb{100}} file is processed only $16\%$ faster.
As the file size increases the effect becomes more visible.
The index of {\gb{1}} prefix is constructed $24\%$ faster, and the
{\gb{5}} index can be built $35\%$ faster.
The improvement in construction time then decreases for the full
{\inenwikibig} collection, as the TLB can only maintain a certain
number of hugepage entries, after which address translation becomes
more costly again.

%
%
%
%
%
%

%

%
%
%
%
%

\section{Related Work}
\label{sec-related}

As part of their experimental work authors often make prototype
implementations of their proposed structures available.
Additionally, several experimental studies and public available libraries focusing on
\SDS\ have emerged over time.
The best known is the widely used {\pizzachili}
corpus\footnote{\href{http://pizzachili.dcc.uchile.cl}{http://pizzachili.dcc.uchile.cl}}, which was
released alongside an extensive empirical
evaluation~{\citep{FER:GON:NAV:VEN:2008}}.
The corpus includes a collection of reference data sets, plus
implementations of several succinct text index structures accessed
via a common interface.
The {\method{libcds}} library is also popular, and provides
implementations of bitvectors and wavelet
trees~{\citep{CLA:NAV:2008}}.
It has recently been succeeded by
{\method{libcds\oldstylenums{2}}}\footnote{\href{https://github.com/fclaude/libcds2}{https://github.com/fclaude/libcds2}}.
{\citet{VIG:2008}} provides bitvector implementations
supporting {\rankop} and {\selectop} in the {\method{Sux}}
library\footnote{\href{http://sux.di.unimi.it}{http://sux.di.unimi.it}}. The
Java version of \method{Sux} also implements minimal
(monotone) perfect hash functions.
Recently {\citet{GRO:OTT:2013}} presented the {\method{succinct}}
library\footnote{\href{https://github.com/ot/succinct}{https://github.com/ot/succinct}} which
consists of bitvectors, succinct tries and a {\RMQ} structure;
all structures can be memory mapped.

Compared to these other implementations, {\sdsl} version 2 has a
number of distinctive features:
it operates on both character and words inputs;
it is optimized for large-scale input (including support for
hugepages);
it provides coverage of a wide range of alternative structures, that
can be composed and substituted in different ways; and it offers
dynamic visualizations that allow detailed space evaluations to be
undertaken.
Finally, fully automated tests and benchmarks are also included, and
can be used by other researchers in the future to check the correctness and
performance of further alternative implementations of the various modules.

%
%

%
%
%
%
%
%
%
%

\section{Conclusion}
\label{sec-conclusion}

We have explored the benefits that flow when modular and composable
implementations of succinct data structure building blocks are
available, and have showed that efficiency at all levels of the
{\SDS} hierarchy can be enhanced by careful attention to low-level
detail, and to the provision of precisely-defined interfaces.
As a part of that demonstration, we introduced the open source
{\sdsl} library, which contains efficient implementations of many
{\SDS}.
The library is structured to facilitate flexible prototyping of new
high-level structures, and because it offers a range of options at
each level of the data structure hierarchy, allows rapid exploration
of implementation alternatives.
The library is also robust in terms of scale, handling input
sequences of arbitrary length over arbitrary alphabets.
In addition, we have demonstrated that the use of hugepages can have
a notable effect on execution times of large-scale {\SDS}; and shown
that the advanced visualization features of the library provide
important insights into the time and space requirements of {\SDS}.

\paragraph*{Acknowledgment.}

The authors were supported by the Australian
Research Council and by the Deutsche
Forschungsgemeinschaft.

\paragraph*{Software \& Experiments.}
The library code, test suite, benchmarks, and a tutorial, are
publicly available at {\href{https://github.com/simongog/sdsl-lite}%
 {https://github.com/simongog/sdsl-lite}}.

\begin{footnotesize}
\bibliographystyle{abbrvnat}
\bibsep 0.5mm
\bibliography{local}
\end{footnotesize}

\end{document}